\documentclass[fleqn,12pt]{article}
\usepackage{epsfig}
\begin{document}
\textheight 22cm
\textwidth 15cm
\noindent
{\Large \bf Predicting PDF tails in systems with logarithmic non-linearity}
\newline
\newline
Johan Anderson\footnote{anderson.johan@gmail.com} and Eun-jin Kim$^{\star}$
\newline
Max-Planck-Institut f\"{u}r Plasmaphysik, IPP-Euratom Association, Teilinstitut Greifswald, 17491 Greifswald, Germany
\newline
$\star$ University of Sheffield, Department of Applied Mathematics, Hicks Building, Hounsfield Road, Sheffield, S3 7RH, UK
\newline
\newline
\begin{abstract}
\noindent
The probability density function (PDF) of flux $R$ is computed in systems with logarithmic non-linearity using a model non-linear dynamical equation. The PDF tails of the first moment flux are analytically predicted to be power law. These PDF tails are shown to be broader than a Gaussian distribution and are a manifestation of intermittency caused by short lived coherent structures (instantons).
\end{abstract}
\newpage
\renewcommand{\thesection}{\Roman{section}}
\section{Introduction}
One of the remaining challenges in classical physics is to understand the nature of turbulence. In particular, gaining information of turbulent systems by experimental determination of probability density functions (PDFs) of flux is very common. While traditional mean field theory is built on Gaussian statistics (such as transport coefficients), avalanche-like events cause the deviation of the PDFs from a Gaussian prediction. In particular, PDF tails due to rare events of large amplitude are often found to be substantially different from Gaussian although PDF centers tend to be Gaussian. {For example the the scaling of PDF tails of fluxes in tokamaks is often found to be exponential~\cite{a24}-~\cite{a22}.} In this paper we present a statistical model of intermittency by computing PDF tails. Our key method is based on the idea that coherent structures such as vortices, streamers and blobs cause intermittency in the PDF tails. Significant flux can be mediated by coherent structures through the formation of avalanche like events of large amplitude, as indicated by recent numerical simulations and experiments~\cite{a10}-~\cite{a20}. 

Specifically, the theoretical method used here to compute the PDF tails is a non-perturbative technique, the so-called the instanton method~\cite{a25}-~\cite{a36}. The instanton method has been adopted from quantum field theory and then modified to classical statistical physics for Burgers turbulence~\cite{a25}-~\cite{a26} and in a model by Kraichnan~\cite{a37}. Note that there is another method determining the PDF tails, namely to compute the Fokker-Planck equation for the PDF~\cite{a36},~\cite{a33}.

In previous papers it has been shown using the instanton method that the PDF tails of momentum flux and heat flux in plasma turbulence are significantly enhanced over the Gaussian prediction~\cite{a27}-~\cite{a30}. In particular, a novel explanation for the exponential scaling of momentum flux $R$ of the form $\sim \exp\{- c (R/R_0)^{3/2}\}$ found in experiments on CSDX at UCSD has been provided~\cite{a21}. It was also shown that shear flows can significantly reduce the PDF tails of Reynolds stress and zonal flow formation~\cite{a31}-~\cite{a34}. In a cubic non-linear model of shear flows the PDF tails were found to be $\sim \exp\{- c (R/R_0)^{4}\}$ and thus fall off much faster than a Gaussian, which was confirmed by numerical simulations~\cite{a33}.

Current models fall short in predicting or interpreting experiments and simulations of an inherently strong intermittent nature. In some cases this involves models that have exponential or even logarithmic non-linear interaction. For example exponential non-linearities may be found in models of current sheaths~\cite{a38} in magnetically confined plasmas. On the other hand, logarithmic non-linearities are found in a variety of models, including a logarithmic non-linear Schr\"{o}dinger equation (LNLSE) introduced in Refs.~\cite{a39}-~\cite{a40}. The LNLSE is appealing from a mathematical point of view in that it supports solitary wave solutions (Gaussons) while still preserving many simple features of linear equations. Specifically in the LNLSE, non-interacting sub-systems are still separable and there exist a lower energy bound. The LNLSE has also been studied in a stochastic reformulation of quantum mechanics (QM) where to each quantum state there exists a stochastic process determined by a Langevin equation~\cite{a41}. However, in QM the non-linear effects have been found in experiments to be very small with a $|c| \sim 10^{-15} eV$ ( $|c|$ is defined below in Eq. (30))~\cite{a411}-~\cite{a413}. Nevertheless, this type of equation is used in many other areas of applications such as nuclear physics, optics and geophysics; see e.g. ~\cite{a414}-~\cite{a417}. In quantum field theory logarithmic non-linearities naturally appear in super-symmetric field theories and inflation cosmology; a Klein-Gordon equation with a logarithmic potential was studied in Refs.~\cite{a42}-~\cite{a43} and have direct applications of the models in Ref.~\cite{a44}-~\cite{a46}.

In principle there are roughly four different types of non-linear interaction terms for fluctuations: exponential non-linearities; power law non-linearities; logarithmic non-linear terms; and cyclic non-linear terms. However for weak fluctuations all these could easily be transformed into power law non-linearities by Taylor expansion which result in PDF tails of the form $\sim \exp\{- c (R/R_0)^{s}\}$ with $s=(n+1)/m$. Here, n and m are the order of the highest non-linear interaction term and moments for which the PDFs are computed, respectively~\cite{a32}. For arbitrary fluctuations, in the case of exponential non-linear interaction the PDF tails was found to be described by the Gumbel distribution which represents a frequency distribution of the extreme values of the ensemble~\cite{a35}. However it is questionable that a Taylor expansion of the non-linear interaction term would be valid for an instanton driven process and thus a more rigorous study is needed.

The purpose of the present paper is to provide analytical expressions of the PDF tails of flux of first moment variables such as potential, density etc in systems with logarithmic non-linearities, i.e. PDF tails of the field itself. The predicted PDFs are shown to be power law for first moment flux,  which are thus enhanced over Gaussian predictions.

The paper is organized as follows. In Sec. II the model non-linear dynamical equation is presented together with preliminaries of the path-integral formulation for the PDF tails. In Sec III the instanton solutions are calculated and in Sec IV the PDF tails are presented. A discussion of the results and conclusion is given in Sec. V. 

\section{Non-perturbative computation of the PDF tails for logarithmic non-linearity}
 Here we present a statistical theory of PDF tails in systems with a logarithmic non-linearity. We follow the procedure developed earlier~\cite{a25}-~\cite{a36} to compute PDF tails in systems governed by logarithmic non-linear interactions. To elucidate the particular features of the logarithmic non-linear interaction we start by computing the PDF tails of the reduced case, keeping only the logarithmic non-linear interaction ($\ln\phi$) and later determine the PDF tails of the equation with a term $\phi \ln \phi$ proposed in~\cite{a39}-~\cite{a40}. It has been shown earlier that PDF tails are rather insensitive to details of the dynamics and depend only on the dominant non-linear term~\cite{a32}. Thus, we consider a non-linear equation in one spatial dimension describing the time evolution of the variable $\phi$ for fluctuations governed by a logarithmic non-linear interaction
\begin{eqnarray}
\frac{\partial \phi}{\partial t} + c \ln(\phi) - \eta \nabla^2 \phi = f.
\end{eqnarray}
Here the term $c \ln(\phi)$ is the logarithmic non-linear interaction term, $\eta$ is a damping term and $f$ is the forcing. For simplicity, the statistics of the forcing is assumed to be Gaussian with a short correlation time modeled by the delta function as
\begin{eqnarray}
\langle f(x, t) f(x^{\prime}, t^{\prime}) \rangle = \delta(t-t^{\prime})\kappa(x-x^{\prime}),
\end{eqnarray}
where $\langle f \rangle = 0$ and $\kappa(x-x^{\prime}) = \kappa_0 e^{(x-x^{\prime})/L^2}$.
The angular brackets denote the average over the statistics of the forcing $f$. The delta correlation in time was chosen for the simplicity of the analysis. In the case of a finite correlation time non-local integral equations in time are needed. 

We calculate the PDF tails of first moment flux $M(\phi)$ (density, potential) by using the instanton method. In general, $M(\phi)$ is the $m$ multiple product of $\phi$ ($m$th moment) and can be denoted $M(\phi) = P_m (\phi)$. In particular, for the first moment $P_1 (\phi) = a \phi$, where $a$ is a constant. The probability density function for the potential $\phi$ can be defined as
\begin{eqnarray}
P(R) =  \langle \delta(M(\phi(x=x_0)) - R) \rangle = \int d \lambda e^{i \lambda R} I_{\lambda},
\end{eqnarray}
where 
\begin{eqnarray}
I_{\lambda} = \langle \exp(-i \lambda M(\phi(x=x_0))) \rangle.
\end{eqnarray}
The integrand can then be rewritten in the form of a path-integral as
\begin{eqnarray}
I_{\lambda} = \int \mathcal{D} \phi \mathcal{D} \bar{\phi} e^{-S_{\lambda}}.
\end{eqnarray}
Here $\mathcal{D} \phi \mathcal{D} \bar{\phi}$ is the path-integral measure. The assumption of Gaussian statistics of the forcing is used to formally express the PDF tails in terms of a path integral~\cite{a36}. 

\section{Instanton (saddle-point) solutions for a system with logarithmic non-linearity}
The path integral in Eq. (5) will be computed by the saddle point method to evaluate the effective action  $S_{\lambda}$ by requiring that the variational first derivatives vanish for the optimum path (among all possible paths). The saddle-point solution of the dynamical variable $\phi(x,t)$ of the form $\phi(x,t) = F(t) v(x)$ is called an instanton if $F(t) = 0$ at $t=-\infty$ and $F(t) \neq 0$ at $t=0$. The optimum path is associated with the creation of a short lived coherent structure (instanton). Thus, the bursty event can be associated with the creation of a coherent structure. Note that, the function $v(x)$ here represents the spatial form of the coherent structure. By using the new variable $\phi = v (1+F)$, we express the action in Eq. (5) by
\begin{eqnarray}
S_{\lambda} & = & -i \int dt c_1 \bar{F}_1 \left(\dot{F} + c_2 \ln(1+F) + \eta c_5 (F+1) \right) \nonumber \\
& + & \frac{1}{2} c_1 c_4 \int dt \bar{F}_1^2 \nonumber \\
& + & i \lambda c_1 c_3 \int dt (1 + F) \delta(t).
\end{eqnarray}
Here, $\bar{F}_1$ is the time dependent part of conjugate variable $\bar{\phi} = \bar{v} \bar{F}_1$ with coherent structure $\bar{v}$. Note that, the sub-leading order (in $\lambda$) term $\bar{v} \ln v \bar{F}_1$ has been neglected. The coefficients $c_1, c_2$, $c_3$, $c_4$ and $c_5$ are defined as
\begin{eqnarray}
c_1 & = & \int dx \bar{v}(x) v(x), \\
c_1 c_2 & = & \int dx c \bar{v}(x), \\
c_1 c_3 & = & \int dx v(x), \\
c_1 c_4 & = & \int dx dy \bar{v}(x) \bar{v}(y) \kappa(x-y), \\
c_1 c_5 & = & \int dx (\nabla^2 \phi) \bar{\phi}.
\end{eqnarray}
The term $\eta (F+1)$ is due to dissipation. To find the instanton solutions we compute the first variational derivatives of $S_{\lambda}$ and take them to be zero
\begin{eqnarray}
\frac{\delta S_{\lambda}}{\delta \bar{F}_1} & = & -i c_1 \left( \dot{F} + c_2 \ln (1+F) + \eta c_5 (1+F)\right) + c_1 c_4 \bar{F}_1 = 0, \\
\frac{\delta S_{\lambda}}{\delta F} & = & -i c_1 \left( \dot{\bar{F}}_1 + c_2 \frac{\bar{F}_1}{1+F} + \eta c_5 \bar{F}_1 \right) + i \lambda c_1 c_3 \delta(t) = 0.
\end{eqnarray}
The equation of motion (EQM) for $t<0$ is derived by differentiating Eq. (12) and substituting the result in Eq. (13)
\begin{eqnarray}
\frac{i}{c_4} \left( \ddot{F} + c_2 \frac{\dot{F}}{1+F} + \eta c_5 \dot{F}\right) = \dot{\bar{F}}_1,
\end{eqnarray}
which leads to
\begin{eqnarray}
-\ddot{F} + c_2^2 \frac{\ln(1+F)}{1+F} + C_5^2 \eta^2 (1+F) + c_2 c_5 \eta \ln (1+F) + c_2 c_5 \eta= 0.
\end{eqnarray}
We assume that dissipation is small but finite and drop the last three terms in Eq. (15). This is valid if $ \eta^2 \lambda$ and $ \eta \ln F$ are sufficiently small. We integrate Eq. (15) by using $u = dF/dt $ and $ \ddot{F} = u du/dF $ and use separation of variables to find
\begin{eqnarray}
\frac{dF}{dt} = \pm c_2 \ln(1+F).
\end{eqnarray}
The instanton function ($F$) now requires the evaluation of the following logarithmic integral
\begin{eqnarray}
\int_{F_0}^{F_t}\frac{dF}{\ln(1+F)} = \pm c_2 (t - \epsilon_{-}).
\end{eqnarray}

Here, $F_0 = \lim_{\epsilon_{-} \rightarrow 0} F(t=\epsilon_{-})$ is the boundary condition at $t=0$ and $F_t = F(t)$ for $t > t_{\infty}$ where $t_{\infty}$ is some large but finite negative time which is bounded due to dissipation. That is, the effect of damping cannot be neglected for $t < t_{\infty}$ (note $t, t_{\infty} < 0$). The resulting logarithmic integral would yield an instanton only when a finite value of dissipation is incorporated. In practice, it is important to note that to compute PDF tails from the action given in Eq. (6) we need only know the value of the first time derivative of the instanton ($\dot{F}$) and its boundary condition at $t=0$ ($F_0$).

Upon integrating Eq. (13) over $(-\epsilon, \epsilon)$, the boundary condition $\bar{F}_1 (\epsilon) = 0$ gives
\begin{eqnarray}
\bar{F_1}(-\epsilon) = c_3 \lambda. 
\end{eqnarray}
We then use Eq. (12) at $t=0$ to obtain
\begin{eqnarray}
\ln(F_0) = - \frac{i c_3 c_4 }{2 c_2} \lambda.
\end{eqnarray}

\section{PDF tails for the logarithmic non-linear interaction}
Here we compute the $\lambda$ dependence of the action $S_{\lambda} \sim h(\lambda)$, which determines $\lambda$-integral in Eq. (3). We first evaluate the action in the large $\lambda$ limit
\begin{eqnarray}
S_{\lambda} & = & -i \int dt 2 c_1 c_2\bar{F}_1 \ln(1+F) \nonumber \\
& + & \frac{1}{2} c_1 c_4 \int dt \bar{F}_1^2 \nonumber \\
& + & i \lambda c_1 c_3 \int dt F \delta(t) \\
& = & \frac{2 c_1}{c_4} \int dt (\ln(1+F))^2 + i \lambda c_1 c_3 \int dt F \delta(t) \\
& \approx & \frac{2 c_1 c_2 }{c_4} \left( F(0) \ln(F(0)) - F(0) \right) + i \lambda c_1 c_3  F(0).
\end{eqnarray}
The action can only be estimated asymptotically for large $\lambda$ since there is a small constant contribution at $t_{\infty}$. The action $S_{\lambda}$ can be expressed as
\begin{eqnarray}
S_{\lambda} \approx   - e^{- i \alpha \lambda},
\end{eqnarray}
where the parameter $\alpha$ is
\begin{eqnarray}
\alpha & = & \frac{c_3 c_4}{2 c_2}. 
\end{eqnarray}
 The tail of the PDF is found by computing the integral in Eq. (3)
\begin{eqnarray}
P(R) \sim \int d \lambda e^{i \lambda R +  e^{- i \alpha \lambda}}.
\end{eqnarray}
In order to use the saddle point method on the $\lambda$ integral, we let
\begin{eqnarray}
f(\lambda) = i R \lambda +  e^{- i \alpha \lambda},
\end{eqnarray}
and find $\lambda_0$, which gives the extreme of the function $f$ (i.e $f^{\prime} (\lambda_0) = 0$) as
\begin{eqnarray}
i R = i \alpha e^{-i \alpha \lambda_0}.
\end{eqnarray}
By taking the natural logarithm on both sides, we conclude that the resulting saddle point is
\begin{eqnarray}
\lambda_0 \approx i \alpha^{-1} \ln(\alpha^{-1} R).
\end{eqnarray}
Using this into Eq. (25) gives us
\begin{eqnarray}
P(R) & \sim & e^{-\frac{R}{\alpha} \ln \frac{R}{\alpha} +  e^{\ln \frac{R}{\alpha}}} \nonumber \\
& = & e^{-\frac{R}{\alpha} (\ln \frac{R}{\alpha} - 1)}.
\end{eqnarray}
The resulting PDF is thus a power law to leading order when $R$ is large.

Next we will consider the $\phi \ln \phi$ non-linear interaction LNLSE and we will show that the PDF tails of first moment flux are close to a Gaussian distribution. In addition, we will calculate the PDF tails of the 2nd moment ($|\phi|^2$) flux and we will show that these tails are similar to the tails of first moment flux of the $\ln \phi$ model (Eq. 29). Following the method used above we find the relevant dynamical equation to be
\begin{eqnarray}
\dot{\phi} + c \phi \ln |\phi|^2 = f.
\end{eqnarray}
Here $f$ is defined in the same way as above. The action can be written as (neglecting subleading order terms)
\begin{eqnarray}
S_{\lambda} & \approx & -i \int dt c_1 \bar{F}_1 (\dot{F} + 2c F \ln F) + \frac{1}{2} c_1 c_4 \int dt \bar{F}^2_1 \nonumber \\
& + & i \lambda c_1 c_3 \int dt F^k \delta (t).
\end{eqnarray}
Here $F$ is assumed to be positive and $k=1$ the for first moment and $k=2$ for the second moment flux. We first consider the case $k=1$. The first variational derivatives of the action $S_{\lambda}$ give dynamical equations for the instanton function $F$ and its conjugate $\bar{F}_1$ that we use to find a relation between the function $F$ and the time derivative $\dot{F}$ as
\begin{eqnarray}
 \dot{F} = 2 c F \ln F,
\end{eqnarray}
and an equation for the value of $F$ at $t=0$
\begin{eqnarray}
F(0) \ln F(0) = \frac{-i c_3 c_4}{4 c} \lambda.
\end{eqnarray}
The action can now be computed in the large $\lambda$ limit to yield
\begin{eqnarray}
S_{\lambda} \approx -\frac{2 c_1 c_3^2 c_4 \lambda^2}{16 c}(\ln (i \frac{c_3 c_4}{c} \lambda) - 4).
\end{eqnarray}
The PDF tails are now found by integration using the saddle-point method with the saddle point $\lambda_0 = -i R/ (2k_1)$
\begin{eqnarray}
P(R) = \int d\lambda e^{i \lambda R - S_{\lambda}} \sim e^{\frac{R^2}{k_1} (- \frac{1}{4} \ln k_2 R + (\frac{1}{k_1}+\frac{1}{2}))},
\end{eqnarray}
Here $k_1 = c_1 c_3^2 c_4 /(16 c)$ and $k_2 = 1/(c_1 c_3)$. We note that these PDF tails are very close to the tails of a Gaussian distribution. In the case of the 2nd moment flux $|\phi|^2$ ($k=2$) we find
\begin{eqnarray}
P(R) \sim e^{- k_3 R ln R + k_3 R},
\end{eqnarray}
with $k_3 = \frac{2c}{c_3 c_4} + \frac{2 c_1 c}{c_4}$ and $k_4 = \frac{2 c_1 c}{c_4}$. 
The PDF tails of 2nd moment flux in LNLSE are significantly enhanced compared to a Gaussian distribution. The PDF drom the 2nd moment are similar in structure to the PDF tails of the first moment in the $\ln \phi$ model due to the logarithmic non-linearity. 

\section{Discussion and conclusions}
In this paper we have presented the first computation of PDF tails of flux in dynamical system with logarithmic non-linear terms ($\ln \phi$). Specifically, the PDF tails of the first moment flux was shown to be a power law. This result is rather insensitive to the precise form of dynamical equation and is only dependent on the dominant non-linear term~\cite{a32}. The PDF tails deviate significantly from Gaussian distributions signifying that rare events with high amplitude influence the total flux.

We now show explicitly that systems with logarithmic non-linearities have fat PDF tails by plotting [Eq. (29)] and comparing the results to a Gaussian distribution. The shown distributions are normalized as one-sided distributions.
In figure 1, the PDF tails of flux with a logarithmic non-linearity and a Gaussian distribution are compared. In the figure the values of the parameter $\alpha$ are $\alpha = 1.0$ (blue, dashed line); $\alpha = 0.5$ (black, dotted line); $\alpha = 2$ (green, solid line); and a Gaussian with a coefficient of unity (red, dashed-dotted line). For logarithmic non-linearity a stretched tail (a power law) is evident. This suggests that the resulting PDF tails in systems with logarithmic non-linearities are stretched compared with Gaussian distributions and that rare bursty events may be predominant in transport processes. Note that the coefficient $\alpha$ is dependent on the coherent structure and the forcing through the constants $c_2$, $c_3$ and $c_4$.

\begin{figure}
  \includegraphics[height=.3\textheight]{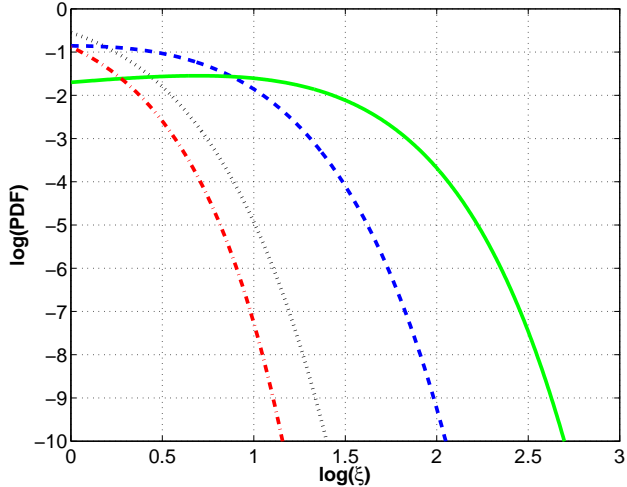}
  \caption{The PDF tails of flux in a system with logarithmic non-linearity and a Gaussian distribution are shown. The values of the parameter $\alpha$ are $\alpha = 1.0$ (blue, dashed line); $\alpha = 0.5$ (black, dotted line); $\alpha = 2$ (green, solid line); and a Gaussian with a coefficient of unity (red, dashed-dotted line). }
\end{figure}

Moreover the usefulness of these PDF tails is limited by the spatial structure of the specific solutions to the non-linear equations. The spatial structure determines the magnitude and sign of the coefficient $\alpha$.

In summary, in the present paper we have investigated a statistical theory of turbulence and intermittency due to coherent structures in dynamical systems with logarithmic non-linearity. The use of coherent structures is motivated by various experimental results where bursty events cause a significant transport which are linked to coherent structures. We have computed the PDF tails of the 1st moment (e.g. density, potential) and 2nd moment ($|\phi|^2$) caused by intermittent coherent structures using the non-perturbative instanton method. The obtained power law PDF tails of first moment flux of $\ln \phi$ model (Eq. (29)) are shown to be significantly enhanced in comparison with a Gaussian distribution. For the LNLSE model with $\phi \ln \phi$ interaction we found PDF tails of the first moment flux close to Gaussian, however the 2nd moment flux were similar to the PDF tails of 1st moment flux of the $\ln \phi$ model. The precise form of the PDF tails in non-linear Klein-Gordon system is however beyond the scope of the present paper and will be studied in future publications. Note that, although Gaussian forcing is used, the non-linearity in the system can give PDF tails that are far from Gaussian. Furthermore, the PDF tails are rather insensitive to the precise details of the dynamical equation and depend only on the dominant non-linear term. This result could provide an important piece of information for interpreting experimental results; the PDF tails derived in here and in previous papers may directly be compared to experiments.

\section{Acknowledgment}
This research was supported by the Engineering and Physical Sciences Research Council (EPSRC) EP/D064317/1.

\end{document}